\documentclass[12pt,preprint]{aastex}

\newcommand{\fuvmag}{\ifmmode{FUV}\else{\it FUV~}}
\newcommand{\nuvmag}{\ifmmode{NUV}\else{\it NUV~}}

\shorttitle{Extinction in M83}
\shortauthors{Boissier et al.}

\def\aha{$A(H\alpha)$}
\def\afuv{$A(FUV)$}
\def\ha{H$\alpha$}
\def\hb{H$\beta$}
\def\b{$\beta$}
\def\bglx{$\beta_{GLX}$}
\def\bi{$\beta_0$}
\def\usfr{M$_{\odot}$ yr$^{-1}$}

\begin{document}


\title{Extinction radial profiles of M83 from GALEX UV imaging} 


\author{Samuel Boissier\altaffilmark{1}, 
Armando Gil de Paz\altaffilmark{1}, 
Barry F. Madore\altaffilmark{1,2},
Alessandro Boselli\altaffilmark{3}, 
V\'eronique Buat\altaffilmark{3} , 
Denis Burgarella\altaffilmark{3},
Peter G. Friedman\altaffilmark{4},
Tom A. Barlow\altaffilmark{4},
Luciana Bianchi\altaffilmark{5}, 
Yong-Ik Byun\altaffilmark{6}, 
Jos\'e Donas\altaffilmark{3},
Karl Forster\altaffilmark{4}, 
Timothy M. Heckman\altaffilmark{7}, 
Patrick N. Jelinsky\altaffilmark{8},
Young-Wook Lee\altaffilmark{6},
Roger F. Malina\altaffilmark{3}, 
D. Christopher Martin\altaffilmark{4},
Bruno Milliard\altaffilmark{3}, 
Patrick Morrissey\altaffilmark{4},
Susan G. Neff\altaffilmark{9}, 
R. Michael Rich\altaffilmark{10},
David Schiminovich\altaffilmark{4},
Oswarld H. W. Siegmund\altaffilmark{8}, 
Todd Small\altaffilmark{4},
Alex S. Szalay\altaffilmark{7}, 
Barry Y Welsh\altaffilmark{8}, and
Ted K. Wyder\altaffilmark{4}
}





\altaffiltext{1}{Observatories of the Carnegie Institution of Washington,
813 Santa Barbara St., Pasadena, CA 91101}
\altaffiltext{2}{NASA/IPAC Extragalactic Database, California Institute
of Technology, Mail Code 100-22, 770 S. Wilson Ave., Pasadena, CA 91125}
\altaffiltext{3}{Laboratoire d'Astrophysique de Marseille, BP 8, Traverse du Siphon, 13376 Marseille Cedex 12, France}
\altaffiltext{4}{California Institute of Technology, MC 405-47, 1200 East California Boulevard, Pasadena, CA 91125}
\altaffiltext{5}{Center for Astrophysical Sciences, The Johns Hopkins University, 3400 N. Charles St., Baltimore, MD 21218}
\altaffiltext{6}{Center for Space Astrophysics, Yonsei University, Seoul 120-749, Korea}
\altaffiltext{7}{Department of Physics and Astronomy, The Johns Hopkins University, Homewood Campus, Baltimore, MD 21218}
\altaffiltext{8}{Space Sciences Laboratory, University of California at Berkeley, 601 Campbell Hall, Berkeley, CA 94720}
\altaffiltext{9}{Laboratory for Astronomy and Solar Physics, NASA Goddard Space Flight Center, Greenbelt, MD 20771}
\altaffiltext{10}{Department of Physics and Astronomy, University of California, Los Angeles, CA 90095}


\begin{abstract}

We use the far-UV (FUV) and near-UV (NUV) images of M83 obtained by GALEX to
compute the radial profile of the UV spectral slope 
in the star
forming disk. We briefly present a model of its chemical evolution
which allows us to obtain 
realistic intrinsic properties of the stellar populations.
Using corollary data, we
also compute the profiles of \ha/\hb{} and of the total IR (TIR) 
to FUV ratio.
Both data and model are used to estimate and compare the extinction
gradients at the FUV wavelength obtained from these various
indicators. We discuss the implications for the determination of the
star formation rate.
\end{abstract}



\section{Introduction}

UV and optical observations of galaxies are in general
affected by dust
extinction. Empirical determinations of the star formation rate in
individual galaxies, modeling of the radial properties and color
gradients of galaxies \citep[e.g.][]{macarthur04}, or
estimations of the ``cosmic star formation rate history''
\citep[e.g.][]{madau98} all need to take into account its 
effects.  
While detailed radiative transfer calculations self-consistently
accounting for the whole SED can be used to estimate this effect
\citep[e.g.][]{popescu2000,popescu2004}, it is usually done by 
applying much simpler standard recipes.

These recipes include the following: (a) The Balmer Decrement
method. This involves a comparison of the observed \ha/\hb{} ratio 
with its predicted value (2.86 for case B
recombination, e.g. Osterbrock 1989).
Although usually used for \ha{} attenuation, its
results can be extrapolated to other wavelengths \citep[e.g.][]{buat02}.
(b) The UV Spectral Slope method.
\citet[][hereafter M99]{meurer99} proposed a 
relation between the UV extinction and the UV spectral slope \b{} (the
continuum spectrum $f_{\lambda}$ between 1300 and 2600 \AA{} having a
shape of $\lambda^{\beta}$) in starbursts. Recently,
\citet[][hereafter K04]{kong04} computed the equivalent relationship
in the GALEX bands.  Finally, (c) the TIR/UV luminosity Ratio method
(see M99).
Radiative transfer models have shown that
this ratio is a robust indicator of the UV extinction as
it does not depend much on either the geometry or the star formation
history \citep{gordon00}.


Each of the above methods has its own limitations. The Balmer Decrement
Method is affected by systematics caused by uncertainties
in the underlying absorption of the stellar populations,
and the fact that it concerns only very massive
stars and may not represent the extinction affecting older stars.
The UV Spectral Slope Method was calibrated for starburst galaxies and may
not apply more generally. 
Moreover, both methods rely on colors determined at similar wavelength, which are quite sensitive
to dust radiative transfer effects (Witt, Thronson \& Capuano 1992, Witt \& Gordon 2000 -hereafter WG2000).
Finally, the TIR/UV method is the most reliable, but
until the advent of the Infrared Space Observatory (ISO)
and now the Spitzer Space Telescope, this method
was limited in its application by the low spatial resolution
of the IRAS data. 
%

%

M83 is an interesting laboratory to test the recipes for
extinction given the profusion of data available in addition to the
GALEX images and for being almost face-on. Models of its stellar
content and chemical evolution can be constrained by these data
and used to predict the intrinsic properties of the stellar
population.  An alternative study combining observations from
GALEX and ISO of M101 is presented in Popescu et al$.$ (this issue).

\section{Observations}

\begin{figure*}
\scalebox{1.2}{\includegraphics*[2cm,7.8cm][15cm,21cm]{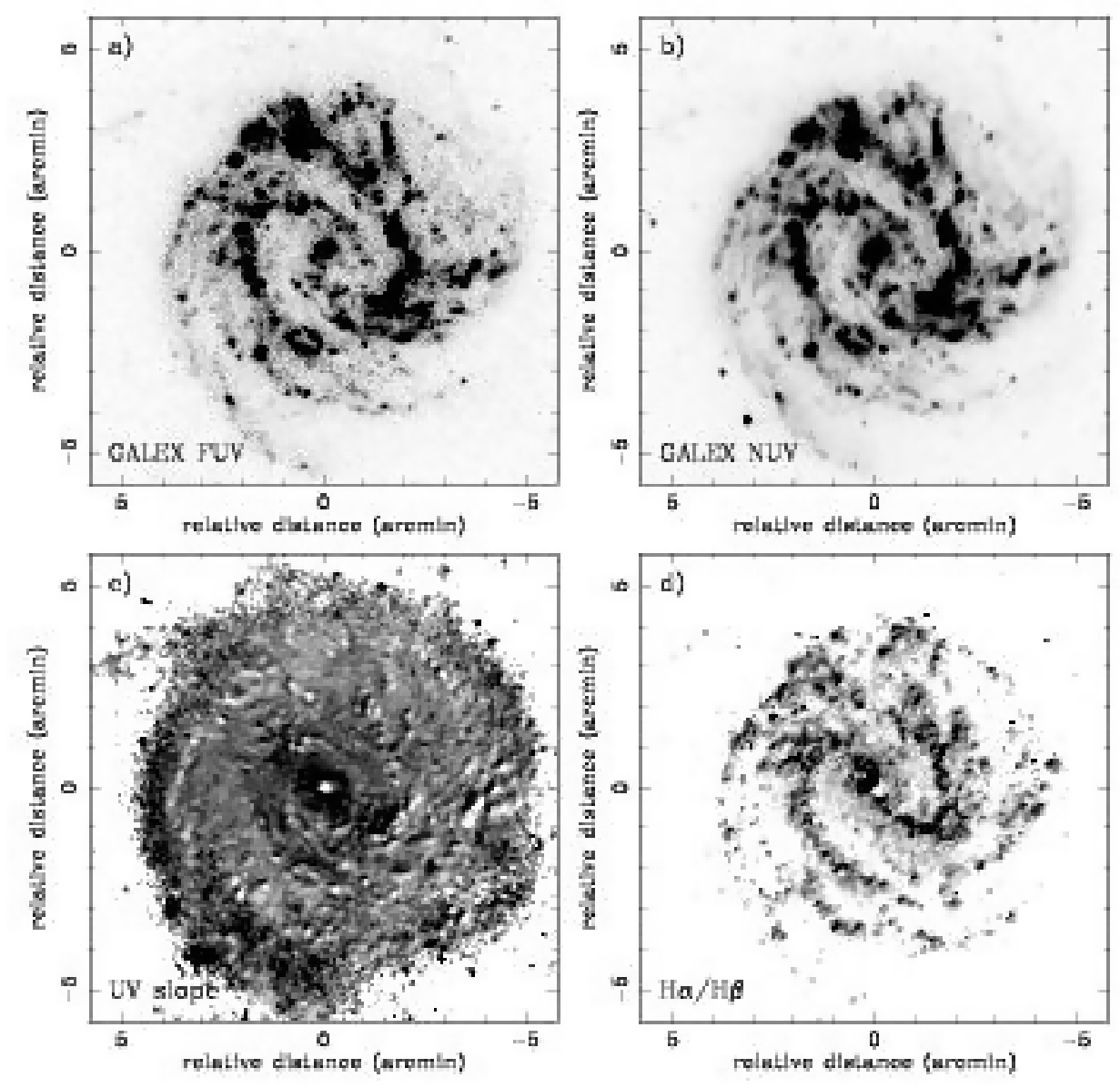}}    
\caption{\emph{a)} FUV image (from the sky background to 22.3 AB mag arcsec$^{-2}$),
\emph{b)}NUV image (from the sky background to 22.7 AB mag arcsec$^{-2}$), 
\emph{c)}UV spectral slope \b, from $-$2.5 (white) to 0.3 (black).
\emph{d)}\ha/\hb{} ratio, from \ha/\hb=2 to 6. White areas in this plot correspond to
regions where either H$\alpha$ or H$\beta$ was below 3 $\sigma_{\mathrm{sky}}$. A
8''$\times$8'' median filter was applied in the case of panels \emph{c} \& \emph{d}.
\label{figMAPS}}
\end{figure*}

GALEX observed M83 for a total of 1352\,s on June 07 2003
simultaneously through its FUV and NUV bands (see Fig$.$ 1). From
these images we derive the FUV and NUV profiles by azimuthally
averaging along ellipses, after masking the stars, and subtracting the
sky. The ellipticity and position angle (0.1 and 80$^o$) were taken
from Kuchinski et al$.$(2000). In this paper, the authors presented
Ultraviolet Imaging Telescope (UIT) observations of M83. The galaxy was
observed by UIT only in the FUV domain, with a slightly better 
resolution ($\sim$3\,arcsec)
than the GALEX one ($\sim$5\,arcsec). 
The surface-brightness sensitivity limit of the GALEX
image is however $\sim$2\,mag deeper. 
Our FUV profile is in good agreement with the one published
in Kuchinski et al$.$(2000) but extend to larger radii.
%
The map and radial profile of the UV spectral slope
$\beta_{GLX}$ (computed in the GALEX bands as
in K04) are shown in Fig$.$  \ref{figMAPS}c \&
\ref{figradial}a, respectively.
We estimate 0.15\,mag to be the 
uncertainty in the
calibration of the GALEX AB surface brightnesses.

We derive the \ha{} and \hb{} profiles
from narrow-band imaging (FWHM$\simeq$60\,\AA) 
in the lines, 
and adjacent continuum images, all 
obtained at the Las Campanas Observatory (Chile) 40-inch
telescope using the 2048$\times$3150 pixels CCD camera. 
The main uncertainty in the \ha{} and especially \hb{} surface brightness
comes from the underlying Balmer absorption that has to be accounted for.
In order to estimate this effect, we use a drift-scanning 
long-slit spectrum of M83 
obtained with WFCCD using the blue grism and slit width of 2\,arcsec at the 100-inch
du Pont telescope also at Las Campanas. Both H$\beta$ absorption and
emission were detected along the central $\pm$4\,arcmin of M83.
The average underlying absorption
at \hb{} is 4.2 \AA{} (with a rms of 1 \AA{} along the slit).
Since the absorption in \ha{} could not be measured, we assumed
it to be the same as for \hb{} (e.g. Mccall et al$.$ 1985).
The adopted value is close to the absorption obtained with the 
stellar population model described in next section 
(4 and 5.5 \AA{} on average at  \ha{} and \hb{} respectively). 
%
We also use the spectroscopic results to estimate the 
[NII]$\lambda\lambda$6548,6584\AA\AA\ contamination and correct 
\ha{} accordingly using
[NII]/\ha=0.45.


Profiles were computed from the 60 and 100 $\mu$m images
obtained by a IRAS-hires request \citep{rice93}.
The 60 $\mu$m image was convolved with a gaussian to obtain a resolution
similar to the 100 $\mu$m image ($\sim$ 90 arcsec). The two profiles
were then combined to compute the TIR 
surface brightness profile, following \citet{dale01}.
The resolution of the FUV images was  
degraded in a similar way 
before computing the TIR/FUV ratios profiles.

\section{Models of M83}

We use models for the chemical and spectro-photometric
evolution of M83 based on \citet{boissier00} (BP2000), with an
updated star formation law (depending on the angular velocity) 
determined in \citet{boissier03}. To compute the SFR, we adopt the 
rotation curve of \citet{crosthwaite02}.
These models, however, cannot reproduce the observed 
gaseous profile  \citep[taken from][]{crosthwaite02,lundgren04}. This is 
most likely due
to the fact that the models do not include radial inflows. There are 
reasons why such a flow might be present in M83: 
the strong bar could induce radial motion
and the central starburst needs to be supplied with fresh gas.
Interaction with NGC~5253 may be responsible
for these peculiarities \citep{vandenbergh80}.
We introduce in the models of BP2000 radial inflows of various efficiencies
and various starting times.
The best models are chosen as the ones reproducing
the observed dust-free profiles: total gas, and near-infrared surface brightness 
of 2MASS. Several 
models including a flow beginning a few Gyrs ago (1 to 5) produce satisfactory
results. 
In Fig$.$ \ref{fig1}a, we show the profile of the average intrinsic
spectral UV slope (\bi, short-dashed curve), and the range of values
(shaded area) obtained with these 
best-models.
\footnote{
In the central 50 arcsec of the galaxy, the presence
of the starburst and of a bulge are not taken into account
in the disk model. In the central starburst itself (10 arcsec), 
the UV spectral slope of the model is artificially changed to the canonical 
value $-$2.1 used for the intrinsic slope of a starburst \citep{calzetti00}.
More details on the models will be given in a forthcoming paper}


\section{Extinction maps and profiles}

We now consider the  ``classical'' extinction estimators used in 
extra-galactic studies:
(a) The \aha{} extinction obtained from the \ha/\hb{} ratio (after
correction for the Balmer underlying absorption)
can be extrapolated to the FUV wavelength.
We obtain \afuv=1.4 \aha{} by
assuming that the color excess of 
the stellar continuum is 0.44  
times the gaseous emission 
one (see Buat et al$.$ 2002 and references within), 
and a Galactic extinction curve (for which \aha/$E(H\beta-H\alpha)$=2.26). 
(b) It is common to 
apply the method of M99 in
galaxies with nuclear starbursts like M83.
We compute a profile following this assumption 
(\b-starburst, Fig$.$ \ref{fig1}b), adopting the TIR/FUV-$\beta_{GLX}$ 
relation computed by K04.
(c) Finally, we compute the extinction \afuv{} from the TIR/FUV ratio
following Buat et al$.$ (this issue).
The result is presented in Fig$.$ \ref{fig1}b. The absence of features in this
gradient is due to the low resolution infra-red IRAS observations.
To achieve a better spatial resolution taking advantage of the TIR/FUV
calibration for \afuv, we perform a fit (outside the 50 central arcsec
of the bulge) providing \afuv=0.86+0.91 (\bglx$-\beta_0$) 
(dotted lines in Fig$.$ \ref{figWG}, shifted to account 
for the average value of $\beta_0$). We use this result to
compute a new extinction profile (\b-models in Fig.2, middle panel)
from the \bglx{} profile.

The extinction maps (Fig$.$ 1) and the profiles (Fig$.$ 2) 
show that the extinction is
relatively low in the very center (see the hole at the center of the 
\bglx-map, Fig$.$ \ref{figMAPS}c), assuming the low values of \bglx{} in this
region indicate low extinction. The extinction (indicated by
the various tracers) becomes rapidly high in a circumnuclear region 
and progressively decreases as we move into the disk of the 
galaxy
%
The various indicators show some common structure: the spectral slope
\bglx{} (and the extinction profiles derived from it), as well as the 
Balmer decrement, show several correlated peaks. One
of the most remarkable is located 100 arcsec from the center. 
Inspection of the 
images (Fig$.$ 1) show that this radius corresponds to the position of 
spiral arms where both the star formation rate
and the extinction are enhanced with respect to average regions of the
galaxy.
The \b-starburst extinction gives larger values compared to 
other indicators, i.e$.$ M83 
shows lower extinctions (computed
from the TIR/FUV ratio) than starburst galaxies for the same value of
$\beta$. This is in agreement with the deviations with respect to the
starburst case already obtained in the LMC (Bell et al$.$ 2002)
or in less active galaxies (K04).

We used earlier a purely empirical relation between \afuv{} 
and (\bglx-\bi) in M83. Because of the many uncertainties
(geometry, \bi...), we prefer this to theoretical 
prescription.  
In Fig$.$ 3a, we compare our data outside the bulge
with two models of WG2000 (adopting for \bi{} 
the value of our best model)
to illustrate the dependence on the geometry .
We also  compare our results with the model of K04 for
the birthrate parameter b obtained in our model of M83 (log b$\simeq$ 0.2).
We find that it overpredicts A(FUV) by a large amount
for any given value of \bglx. Finally,
our data are compared to other samples: the fit of K04 for
the starbursts of M99 (solid line), and the integrated NUV and TIR
selected galaxies of Buat et al$.$ (this issue). The profile of 
M83 is consistent with the latter, althought with slightly larger
$\beta_{GLX}$.
%
%
Fig$.$ \ref{fig2}b shows the \afuv/(\b$-\beta_0$) ratio (where \afuv{} is
still determined from TIR/FUV ratio) as a function of radius (the dotted
curve
indicates the same profile if the extinction derived from (\b$-$\bi) as discussed
previously is used
instead of the TIR/FUV ratio). 
Outside the bulge,  \afuv/(\b$-\beta_0$) has an average value of 1.9 (this value 
would become 1.2 if the usual \bi=$-$2.1 was adopted).



\section{Implications for the star formation rate}

\begin{figure}
\includegraphics[width=0.5\textwidth]{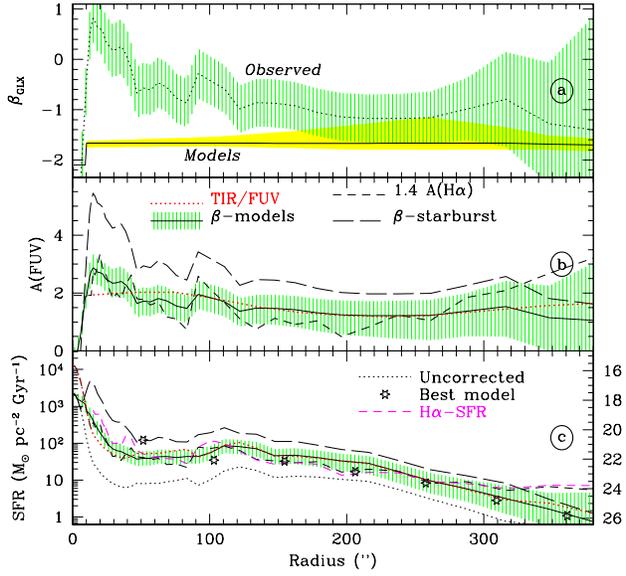}
\caption{
\emph{a)} Radial profile of the observed UV spectral slope
(dotted, derived from the GALEX bands).
The solid curve (shaded area) indicates at each radius the average \bglx{}
(the range of values) obtained with models providing satisfactory profiles. 
\emph{b)} FUV extinction profiles derived from various indicator
(see legend in the figure and details in text).
\emph{c)} FUV-deduced Star Formation Rate profiles 
(surface brightness scale on the right in mag arcsec$^{-2}$), 
uncorrected for extinction (dotted), and with the extinction corrections
given in panel b. The stars show the current SFR obtained with the model.
We also show the SFR deduced from \ha{} corrected for \aha.
\label{fig1}
\label{figradial}
}
\end{figure}

\begin{figure}
\scalebox{0.48}{\includegraphics*[2.cm,15.5cm][20cm,24.5cm]{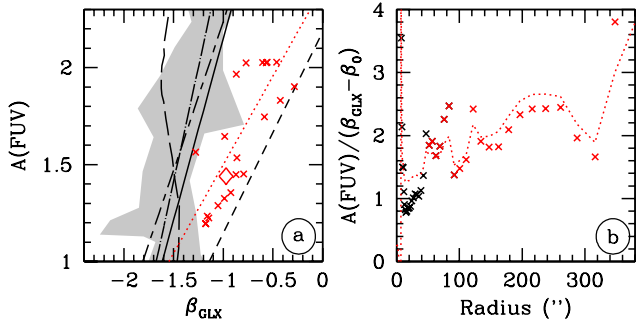}}
\caption{
\emph{a)} The crosses show the FUV extinction (\afuv{} from the TIR/FUV ratio) 
as a function
of \bglx{} for our data out of the bulge (i.e$.$ the central 50 arcsec). The dotted line is
least-square fit.
The shaded area includes the integrated values found by Buat et al$.$ (this
issue).
The short-long dashed line is the relation found for starbursts in 
K04.
The solid curve is obtained with the model of K04, for
a present to past averaged star formation ($b$) such that log $b$=0.2,
typical value obtained with our model of M83.
The short and long dashed curves shows the relation expected for
different geometrical models of WG2000 (respectively 
shell-homogeneous and dusty-clumpy).
\emph{b)} FUV extinction to (\b$-$\bi) ratio as a function of radius (crosses). 
The dotted curve is obtained using the previous fit instead of \afuv. 
\label{figWG}
\label{fig2}}
\end{figure}

The FUV surface brightness profile can be transformed into a star
formation rate profile using the standard conversion factor of
\citet{kenni98}: SFR (\usfr) = 1.4 $\times$ 10$^{-28}$ L$_{UV}$ 
(erg s$^{-1}$ Hz$^{-1}$).  Regardless of the
uncertainties affecting this number (see Kennicutt 1998), 
the FUV profile must be corrected for extinction to obtain a reliable SFR.
Fig$.$ \ref{fig1}c shows the SFR profile obtained without extinction 
correction (lower dotted line), and with various corrections
proposed in the previous section\footnote{We also show the SFR
profile deduced from the \ha{} surface brightness profile, corrected
from \aha, and it is very similar to the SFR deduced from the
FUV. This fact indicates that the current azimuthally averaged SFR is
not very different from the SFR averaged over the last $10^8$ years.}.  
We find that the SFR profile of our best model (stars in Fig$.$ 2c)
is in good agreement with the one deduced from the observations
with the TIR/FUV correction. This suggests that the 
star formation law used in our evolutionary models is 
quite realistic.




The integrated star formation rate also depends on the extinction
correction. 
Owing to multi-wavelength resolved data, 
we can compute the SFR of M83 inside 300
arcsec according to various assumptions.
With no extinction correction, the derived star formation rate is 1
\usfr. Assuming the whole galaxy behaves like a starburst
(\b-starburst), we obtain 10.1 \usfr.  This assumption is probably
over-estimating the star formation rate as the starburst is located
only in the center. The extinction based on the Balmer decrement gives
a SFR of 3.3 \usfr{} and the one based on a combination of the TIR/FUV ratio and \bglx{}
profiles provides 3.6 \usfr{} (a similar value of 4 \usfr{} is
obtained from the low-resolution TIR/FUV extinction profile).


To summarize, we used GALEX observations of M83 in combination with
other data to compare several extinction indicators (the UV Spectral 
Slope method
giving poor results), and to estimate the effect of extinction on the 
determination of the SFR. Differences of a factor up 
to 2.5 are obtained with the various methods. Moreover, we 
determined an extinction-free SFR profile allowing to check
the validity of the SFR law used in evolutionary models.
A similar analysis will be applied to a larger number of
galaxies observed by GALEX in the future.

GALEX (Galaxy Evolution Explorer) is a NASA Small Explorer, launched
in April 2003. We gratefully acknowledge NASA's support for
construction, operation, and science analysis for the GALEX mission,
developed in cooperation with the Centre National d'Etudes Spatiales
of France and the Korean Ministry of Science and Technology.  We thank
C. Popescu and R. Tuffs for their comments. S.B. thanks the CNES
for its financial support.




\clearpage

\end{document}